\documentclass[aps,prl,twocolumn,showpacs,amsmath,amssymb,floatfix,superscriptaddress]{revtex4-1}
\usepackage{graphicx}
\usepackage{dcolumn}
\usepackage{bm}
\usepackage{color}
\usepackage{graphicx}
\usepackage{epstopdf}
\usepackage{epsfig}
\usepackage[colorlinks=true, citecolor=blue, anchorcolor=green]{hyperref}

\makeatletter
\begin{document}
\include{graphic}
\preprint{APS/123-QED}

\title{Time-Varying Strong Coupling and Its Induced Time Diffraction of Magnon Modes}

\author{Jinwei~Rao}
\affiliation{School of Physical Science and Technology, ShanghaiTech University, Shanghai 201210, China}
\affiliation{School of Physics, Shandong University, No. 27 Shandanan Road, Jinan, 250100 China}
\affiliation{State Key Laboratory of Infrared Physics, Shanghai Institute of Technical Physics, Chinese Academy of Sciences, Shanghai 200083, China}

\author{Yi-Pu~Wang}
\affiliation{Zhejiang Key Laboratory of Micro-Nano Quantum Chips and Quantum Control, School of Physics, Zhejiang University, Hangzhou 310027, China}

\author{Zhijian~Chen}
\affiliation{School of Physical Science and Technology, ShanghaiTech University, Shanghai 201210, China}

\author{Bimu~Yao} \email{yaobimu@mail.sitp.ac.cn;}
\affiliation{State Key Laboratory of Infrared Physics, Shanghai Institute of Technical Physics, Chinese Academy of Sciences, Shanghai 200083, China}
\affiliation{School of Physical Science and Technology, ShanghaiTech University, Shanghai 201210, China}

\author{Kaixin~Zhao}
\affiliation{School of Physical Science and Technology, ShanghaiTech University, Shanghai 201210, China}

\author{Chunke~Wei}
\affiliation{State Key Laboratory of Infrared Physics, Shanghai Institute of Technical Physics, Chinese Academy of Sciences, Shanghai 200083, China}

\author{Congyi~Wang}
\affiliation{School of Physical Science and Technology, ShanghaiTech University, Shanghai 201210, China}

\author{Runze~Li} 
\affiliation{School of Physical Science and Technology, ShanghaiTech University, Shanghai 201210, China}

\author{L. H.~Bai} \email{lhbai@sdu.edu.cn;}
\affiliation{School of Physics, Shandong University, No. 27 Shandanan Road, Jinan, 250100 China}

\author{Wei~Lu}\email{luwei@shanghaitech.edu.cn;}
\affiliation{State Key Laboratory of Infrared Physics, Shanghai Institute of Technical Physics, Chinese Academy of Sciences, Shanghai 200083, China}
\affiliation{School of Physical Science and Technology, ShanghaiTech University, Shanghai 201210, China}

\begin{abstract}

Time-varying media break the temporal translation symmetry of wave propagation in materials, enabling advanced wave manipulations. However, this novel phenomenon has been rarely explored in magnonic systems due to the significant challenge of achieving a sudden and prominent change in magnon dispersion within materials. Here, we construct a time-varying strong coupling between two magnon modes, and observe a change in the beats of Rabi-like oscillations near the pulse edges. Using a frequency-comb spectroscopy technique developed in this work, we characterize the frequency conversion of magnon modes induced by the time-varying strong-coupling effect. Moreover, we construct time slits with adjacent time interfaces and demonstrate, for the first time, the double-slit time diffraction of magnon modes, analogous to the well-known Young's double-slit experiment. These findings rely solely on the time-varying strong magnon coupling, independent of device reconfiguration. Our results open avenues for applications such as all-magnetic mixers or on-chip GHz sources.

\end{abstract}
\maketitle

\begin{figure*} [htbp]
\begin{center}
\epsfig{file=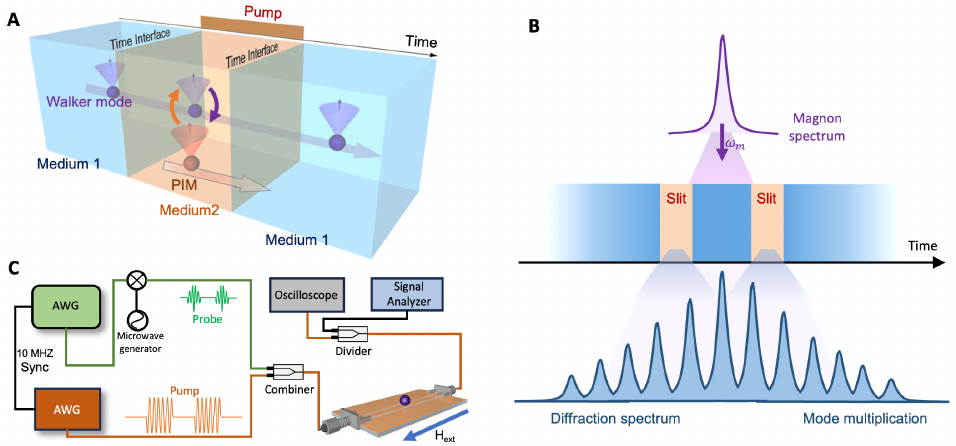, width=16cm} \caption{\textbf{Constructing time slits for double-slit diffraction of magnon modes.} (A) A schematic diagram of the time interfaces of magnon modes and the time evolution of the coupled PIM-WM system driven by a pump pulse. (B) An incident magnon mode (i.e., standing spin wave) at $\omega_m$ diffracted by two time slits, resulting in the formation of interference fringes in spectrum. (C) Experiment set-up of the FCS measurement, which is used to characterize time diffraction of magnon modes. }
\label{TD}
\end{center}
\end{figure*}

Time-varying media, materials whose physical properties change over time, have attracted much attention for their unique advantages in wave manipulation. Unlike in classical wave physics, where wave propagation is controlled by disrupting the spatial translation symmetry of the host medium using reflectors, lenses, or apertures, in the time domain, wave manipulation is achieved by breaking the temporal translation symmetry. This leads to emerging phenomena such as time reflection \cite{moussa2023observation, bar2024time, jones2024time}, time refraction \cite{Schultheiss, dong2024quantum, zhou2020broadband, dikopoltsev2022light}, and time diffraction \cite{Moshinsky, Szriftgiser, tirole2023double}. These nontrivial wave manipulations based on time-varying media are not limited by the dimensions of the medium and can operate on subperiod scales, offering distinct merits. Consequently, they are being studied in depth across various disciplines \cite{Cronin2009,nagulu2020non}, including atomic, photonic, electronic, and phononic research. So far, many novel functions, such as new forms of gain \cite{galiffi2019broadband}, multimode light shaping \cite{chamanara2019simultaneous}, and ultrafast switching \cite{shcherbakov2015ultrafast}, have been successfully realized.

Magnons, as the energy quanta of spin waves in materials, can carry information in insulators and holds its potential as a complementary technology to CMOS \cite{pirro2021advances,zhang2015magnon,wang2021magnonic,rao2019analogue}. Thus, they possess promise for applications in wave-based computing with low dissipation and in the construction of quantum hybrid systems as quantum transducers \cite{tabuchi2015coherent,wang2023qua}. In recent years, we have witnessed the flourishing of time-varying photonics; however, its counterpart---the field of time-varying magnonics that studies the spin wave propagation in time-varying media---remains unexplored. The main challenge lies in the lack of efficient methods for rapidly modulating the magnon dispersion in magnetic materials. Due to the large magnetic moment of most magnetic materials, rapidly modulating the dispersion of normal magnon modes \cite{xu2020floquet, schultheiss2021time} requires a pulse field with an intensity of at least millitesla. This leads to significant energy consumption and Joule heating produced by pulse currents, which directly undermines the potential of magnons as low-dissipation information carriers and building blocks in quantum hybrid systems.

The recently discovered pump-induced magnon mode (PIM) \cite{rao2023unveiling,wang2024enhancement} provides an ideal platform for developing time-varying magnonic systems. Unlike normal magnon modes, the magnetic moment of a PIM is significantly smaller, allowing its momentum to respond readily to weak microwave magnetic fields (around $\sim\rm{nT}$). In this work, we demonstrate the time-varying strong coupling of magnon modes by introducing the PIM, subsequently observing the time diffraction of magnon modes. In experiment, short microwave pump pulses are used to induce the formation and waning of the PIM in a yttrium iron garnet (YIG) sphere. The coupling between the PIM and the Walker mode (WM) in the YIG sphere \cite{walker1957magnetostatic, dillon1957ferrimagnetic, gloppe2019resonant} varies over time, characterized by an increase in coupling strength at the leading edge of the pump pulse and a decrease at the trailing edge. The spectral evolution of this time-varying strong coupling is characterized using our time-resolved frequency comb spectroscopy (FCS) developed in this work. This technique effectively overcomes the challenge of short-time spectral sampling and eliminates cross-talk from microwave pulses. By leveraging this technique, we observe the sudden change in magnon dispersion induced by the time-varying strong PIM-WM coupling. Such sudden change defines two distinct magnetization states of the YIG, thereby forming a time interface for spin waves (or magnon modes) (Fig. \ref{TD} A). Two adjacent time interfaces create a ``time slit", leading to time diffraction and spectral broadening of the magnon mode (Fig. \ref{TD} B). This feature allows for precise control over the mode multiplication by constructing a double-slit interferometer for magnon modes in the time domain. Similar to Young's double-slit experiment, the frequency spacing of the multiplied magnon modes is inversely proportional to the slit separation, demonstrating the double-slit time diffraction of magnon modes (Fig. \ref{TD} B). These findings pave the way for nonlinear spintronic applications, such as on-chip spin wave sources and all-magnetic mixers.

\begin{figure} [htbp]
\begin{center}
\epsfig{file=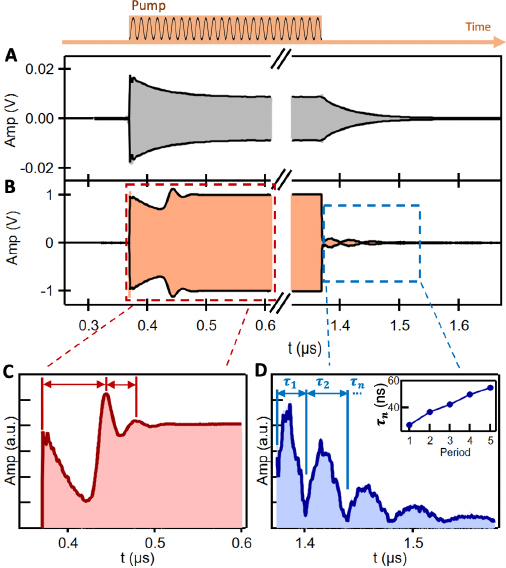,width=8.8cm} \caption{\textbf{Time-varying strong PIM-WM coupling.}  (A) Measured transmission waveform at low pump power. (B) Rabi-like oscillations occurring at the leading and trailing edges of the pump pulse, when setting the pump pulse at a high power level. (C) and (D) Zoomed-in views of the oscillations at pulses edges. The beats exhibit opposite variation behaviors at the leading and trailing edges. Inset: Duration of each beat extracted from D. }\label{PandP}
\end{center}
\end{figure}

In our experiment, the YIG sphere is mounted on top of a standard coplanar waveguide and immersed in a static magnetic field (Supplementary Section 2 \cite{SM}). The (2,2,0) WM is fixed at 3.6 GHz. A high-speed waveform generator sends periodic pump pulses to excite the PIM for the dynamic modulation of the PIM-WM coupling. The repetition time of the pump pulses is significantly longer than the magnon lifetime, allowing us to treat the excitation of each pump pulse as an independent event. The transmission signal of the pump pulses is monitored by an oscilloscope to characterize the temporal evolution of magnon modes (indicated by orange lines in Fig. \ref{TD} C). Figures \ref{PandP} A and B show two typical waveforms at different instantaneous pump powers. When the instantaneous power is low ($\sim-20$ dBm), the  effective spin number of the PIM is small, rendering its interaction with the WM negligible. Two exponential damping processes with identical decay time ($\sim$ 0.5 $\mu$s) occur at the leading and trailing edges of the pump signal (Fig. \ref{PandP} A), corresponding to the excitation and relaxation of the WM. However, when we increase the instantaneous pump power to a high level ($\sim 27$ dBm), Rabi-like oscillations \cite{Zhang2014} occur at both the leading and trailing edges of the pump pulse (Fig.\ref{PandP} B). These oscillations reflect the evolution of the PIM-WM coupling during the formation and waning of the PIM. 

Distinct from those Rabi-like oscillations in conventional strong coupled systems, the beat periods in our system gradually decrease at the leading edge and increase at the trailing edge as shown in Figs.\ref{PandP} C and D, respectively. For instance, each beat period $\tau_n$ from Fig. \ref{PandP} D is summarized in the inset, from which we can clearly see the prolongation of the beat period. It suggests a reduction in the PIM-WM coupling strength after switching off the pump. These gradual changes in oscillation beats resemble the time diffraction of particle beams predicted by Moshinsky \cite{Moshinsky}. They can be interpreted as the time diffraction of magnon modes at 'straight edges' \cite{vedad2024precision} created by the pump pulse. The mechanism hinges on the spin number change of the PIM during its formation and waning.

\begin{figure*} [htbp]
\begin{center}
\epsfig{file=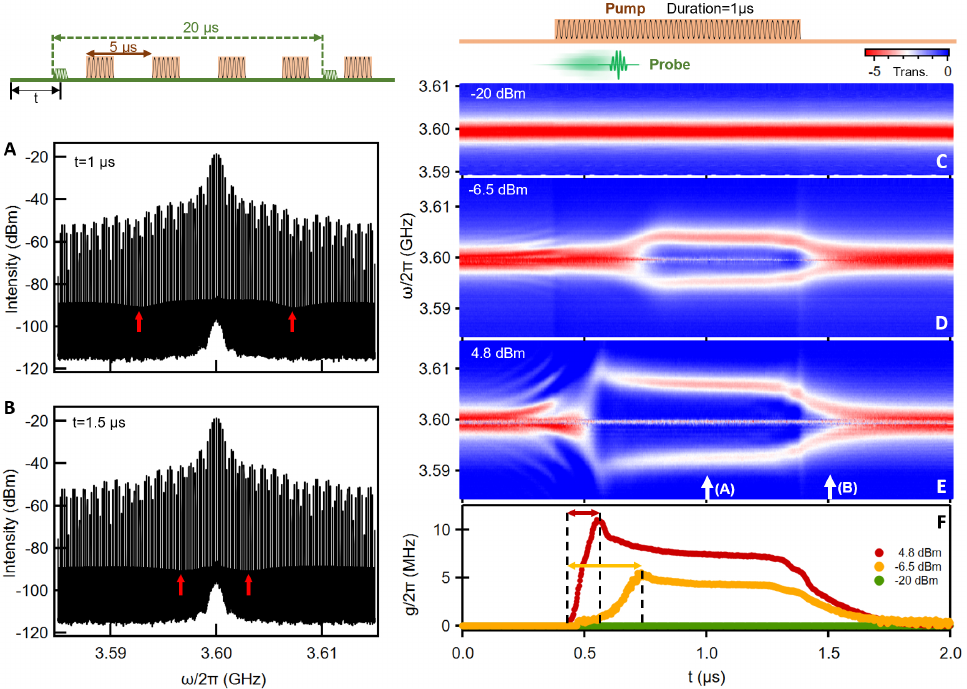,width=16cm} \caption{\textbf{Spectra evolution of the coupled PIM-WM system in time domain.} Up panel: Schematic of pump and probe signals with different repetition periods. (A) Frequency spectrum of the synthesized signal measured at $t=1$ $\mu$s. It is the superposition of two frequency combs that correspond to the pump pulses and the probe pulses. The frequency comb with dense teeth and low intensities ($\sim-85$ dBm) corresponds to the probe pulse. (B) Frequency spectrum of the synthesized signal measured at $t=1.5$ $\mu$s. Red arrows in (A) and (B) indicate the two eigenmodes of the coupled PIM-WM system. (C)-(E) Transmission spectra of the coupled PIM-WM system measured at different delay times, when setting the instantaneous power of the pump signal at -20 dBm, -6.5 dBm and 4.8 dBm, respectively. White arrows in (E) mark the two transmission spectra extracted from (A) and (B). (F) Coupling strength $g$ as a function of $t$.}\label{tdmap}
\end{center}
\end{figure*}

The frequency variation produced by the time diffraction of magnon modes occurs on the nanosecond scale, which is much faster than the acquisition speed of commercial vector network analyzers. To address this challenge, we developed the FCS technique to characterize the magnon spectrum in real time. The experimental set-up is shown in Fig. \ref{TD} (C). The pulse sequence setting is shown in the schematic in Fig. \ref{tdmap}. We combine the periodic pump (orange) and probe (green) pulses and then send the synthesized signal to the YIG sphere. We fix the start time of the pump pules, and define the central time of the probe pulse as $t$. By adjusting $t$, we can stagger the probe pulses from the pump pulses in the time domain, so that the pump-probe delay is changed. The carrier frequencies of the pump and probe pulses can be adjusted to avoid temporal interference between them; in the measurement shown in Fig. \ref{tdmap}, we simply set their carrier frequencies at 3.6 GHz. The periodic pump and probe pulses correspond to two sets of microwave frequency combs, as shown in Fig. \ref{tdmap} A and B. These two spectra are measured at $t=1$ $\mu$s and 1.5 $\mu$s, corresponding to the overlap and non-overlap of the pump and probe pulses, respectively. The comb with sparse and strong teeth corresponds to the pump pulses, while the comb with very dense teeth corresponds to the probe pulses. The probe comb has an intensity much weaker than that of the pump pulses. Therefore, we can neglect the PIM excited by the probe pulse and only use it to measure the time-varying PIM-WM coupling. During the FCS measurement, we adjust the start time of the probe pulses and simultaneously extract the variation of the probe comb envelope in the frequency domain. From the envelop variation (Fig. \ref{tdmap} A and B), we can clearly see two resonant dips marked by red arrows, corresponding to the two eigenmodes of the strongly coupled PIM-WM system. When the pump-probe delay changes, due to the different excitation status of the PIM, the splitting gap between two dips varies, indicating the temporal variation of the coupling strength between the PIM and WM.

Leveraging the FCS technique, we can characterize the time-varying coupling of the coupled PIM-WM system induced by the pump pulse. Figure \ref{tdmap} C-E show the results measured at three different power levels. The duration time of the pump is set as 1 $\mu$s, with its instantaneous power at -20 dBm, -6.5 dBm and 4.8 dBm, respectively. We continuously change the pump-probe delay by adjusting $t$, and extract the transmission spectrum from the probe comb at each $t$. When the pump power is low, PIM excitation is nearly negligible. From the transmission spectra, only a single resonance corresponding to the WM can be observed (Fig. \ref{tdmap} C). Increasing the pump power to -6.5 dBm and repeating the measurement, we can see that the single resonance gradually splits into two resonances after the leading edge of the pump pulse, indicating the formation of the strong PIM-WM coupling. After that, the coupled PIM-WM system evolves into an equilibrium state, where the formation and decay rates of the PIM balance out, resulting in a constant splitting gap. Once the pump pulse ends, this delicate balance is disrupted, leading to a decrease in the PIM-WM coupling and causing the coalescence of two resonant dips. Further increasing the pump power to 4.8 dBm, the formation of the strong PIM-WM coupling after the leading edge becomes much faster than the low power case. The mode splitting rapidly reaches its peak at $t=0.58$ $\mu$s within nanosecond scale. This sudden change in magnon dispersion forms a time interface, leading to the diffraction of magnon modes. Consequently, in addition to the primary WM, sideband resonances emerge near the leading edge of the pump pulse due to time diffraction of magnon modes (Fig. \ref{tdmap} E). 

\begin{figure*} [htbp]
\begin{center}
\epsfig{file=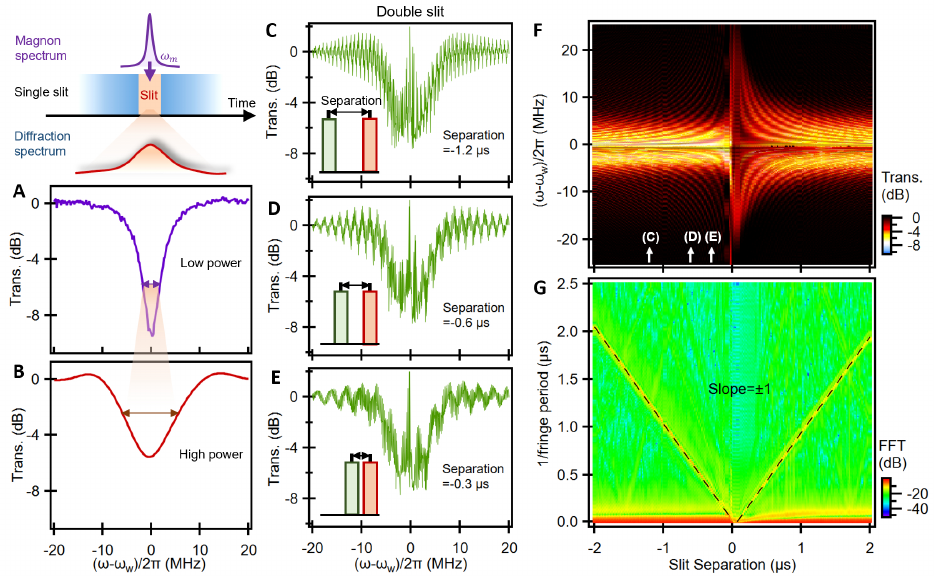,width=15.5cm} \caption{\textbf{Time diffraction of magnon modes.} (A) Up panel: Sketch of a single time slit with pulse duration of 30 ns. Bottom panel: Transmission spectrum measured at low pulse power. In this case, there is no PIM generation and only a WM spectrum is observed. (B) Resonance broadening when the pulse power is high enough to excite PIMs. (C)-(E) Temporal evolution of the transmission spectra, when setting the time separation between two pump pulses at -1.2, -0.6 and -0.3 $\mu$s, respectively. The periodic side-band resonances arising from the double-slit diffraction of magnon modes occur in the transmission spectra. (F) Interferogram of magnon modes as a function of the slit separation, showing mode multiplication with equal frequency spacing. (G) Fringe periods at different slit separation extracted from F, which approximately follow an inverse proportion of the slit separation.}
\label{dsd}
\end{center}
\end{figure*}

The time-varying PIM-WM coupling is governed by a time-dependent Hamiltonian (see supplementary section 1 \cite{SM}). The variation in the coupling strength ($g$) reflects the evolution of the effective spin number of the PIM ($N_p$) over time, which can be phenomenologically described by
\begin{equation}
    \frac{dN_p}{dt}=-\frac{\lambda}{1+\gamma N_p}\left[N_p-\langle\hat{b}^\dagger\hat{b}\rangle\right],
    \label{Var_N}
\end{equation}
\noindent where $\lambda$ and $\gamma$ respectively represent the linear and nonlinear relaxation coefficients of $N_p$. $\hat{b}$ ($\hat{b}^{\dagger}$) represent the annihilation (creation) magnon operators of PIM. When these two processes balance, the coupled PIM-WM system reaches its equilibrium state. After the pump pulse, the waning of the PIM results in the reduction in both $N_p$ and $g$ until they reach zero. The periods with $N_p=0$ and $N_p\neq0$ define two distinct states of the material. As long as the transition between adjacent periods is sharp enough, a time interface is formed.

Fitting the transmission spectra in Fig. \ref{tdmap} C-E by using the expression $S_{21}=1+\kappa_w/[i(\omega-\widetilde{\omega}_w)+g^2/i(\omega-\widetilde{\omega}_u)]$, we extract $g$ at different time. Here, $\kappa_w$, $\widetilde{\omega}_w$ and $\widetilde{\omega}_u$ represent the external damping rate of the WM, the complex frequencies of the WM and PIM, respectively. The results are summarized in Fig. \ref{tdmap} F. The temporal variation of $g$ can be simulated by using Eq. (\ref{Var_N}) (see supplementary section 1 \cite{SM}). This variation consists of three consecutive exponential processes at a high pump power, indicating the formation, stabilization and waning of the coupled PIM-WM system. As the pump power increases, the formation period of the strong PIM-WM coupling is significantly shortened (see Fig.\ref{tdmap} F), leading to a transition in the material state towards a sudden change, i.e., a time interface.

By increasing the pulse intensity while simultaneously lowering the pulse duration, the rapid formation and waning of the PIM create a pair of time interfaces. Between them, a time slit for magnon modes is formed. The single-slit diffraction of magnon modes leads to a broadening of the magnon spectrum \cite{tirole2022single}. The experimental results are shown in Fig. \ref{dsd} A and B. When the pulse intensity (-17.5 dBm) is too weak to excite the PIM for forming the time slit, only the WM mode exists in the transmission spectrum (Fig. \ref{dsd} A). However, as the pulse intensity increases to 21.7 dBm, the time varying PIM-WM coupling enables the time diffraction of magnon modes. A significant broadening of the resonance is observed in the transmission spectrum (Fig. \ref{dsd} B).

Following this idea, we construct two time slits by two short pump pulses (parameters listed in the supplementary section 3 \cite{SM}) and realize the double-slit time diffraction of magnon modes. The WM ($\omega_w$) is tuned to 3~GHz to enhance the magnon coupling \cite{rao2023unveiling}. The carrier frequencies of these two pulses are set at $\omega_w$ with a slight detuning of 25~kHz to avoid their temporal interference. The transmission spectra are extracted from the frequency combs of the first pulse (colored green in Fig. \ref{dsd}), as we change the slit separation between two pulses. Figure \ref{dsd} C-E show three typical transmission spectra measured at the slit separation of -1.2, -0.6, and -0.3 $\mu$s where the negative sign indicates the first pulse occurs earlier. In addition to a broad resonance centered at zero detuning ($\omega-\omega_w=0$), we can see mode multiplications with equal spacing symmetrically distributed with respect to zero detuning. These periodic modes are the interference fringes arising from the double-slit time diffraction of magnon modes. As the slit separation decreases, the period of these fringes increases. The continuous transition of these fringes are plotted in Fig. \ref{dsd} F. The interferogram is symmetric with respect to the zero slit separation, where the period of the fringes reaches the maximum. Using the Fourier transform, we can extract the reciprocal of the fringe period at each slit separation. The results are shown in Fig. \ref{dsd} G, from which we can see two straight lines with slopes of $\pm1$ (see supplementary section 4 \cite{SM}). 

To further validate our conclusion regarding the time diffraction of magnon modes, we performed two additional measurements to rule out other possible origins of the observed fringes (Supplementary Section 3 \cite{SM}). First, we repeat the measurement shown in Fig. \ref{dsd} F at zero magnetic field, where no magnon mode is present, resulting in the disappearance of fringes. Next, we tune the WM frequency to 4.6 GHz by increasing the magnetic field, a frequency too high for the microwave to excite the PIM \cite{rao2023unveiling}.  This ensures a time-independent magnon dispersion during the pulse, and the results have no fringes.  These results reinforce our conclusion that the fringes in Fig. \ref{dsd} F arise from the time diffraction of magnon modes.

In conclusion, we have demonstrated the time diffraction of magnon modes through the rapid modulation of the strong PIM-WM coupling using microwave pulses. This phenomenon is rooted in the transient formation and waning of PIM, which can be utilized to create time slits for magnon mode oscillation. To elucidate this effect, we developed a FCS technique that enables real-time characterization of the magnon mode variation. By leveraging this technique, we observed the interferogram of magnon modes, analogous to the Young's double-slit experiment. This discovery paves the way for the efficient multiplication of magnon modes, advancing the rapid and programmable control of magnon dynamics using tailored pulse sequences. The time-varying control with coherence further holds the promise for spin wave conversion and amplification. Additionally, the time-resolved frequency spectrum technique we developed provides a valuable tool for studying dynamic systems in the microwave range. Our findings are not only significant for the field of magnonics but also have the potential to advance information processing and detection technologies.

\subsection*{Data availability}

The data that support the findings of this study are available from corresponding authors upon reasonable request.

\subsection*{Acknowledge}
This work has been supported by the Strategic Priority Research Program of the Chinese Academy of Sciences (Grant No.XDB0580000), the National Natural Science Foundation of China (Grant Nos. 12204306, 12122413, 11991063, 12474120 and 12227901), STCSM (Nos. 23JC1404100, and 22JC1403300), and the Strategic Priority Research Program of CAS (No.XDB43010200), the National Key R$\&$D Program of China (Nos.2022YFA1404603 and 2022YFA1604400), the Shandong Provincial Natural Science Foundation, China (Grant No. ZR2024YQ001), Qilu Young Scholar Programs of Shandong University. Yi-Pu Wang is supported by the National Key Research and Development Program of China (No.~2022YFA1405200 and No.~2023YFA1406703), National Natural Science Foundation of China (No.~$92265202$ and No.~$12174329$)

\subsection*{Competing interests:} The authors declare no competing interests.

\subsection*{Correspondence} Correspondence and requests for materials should be addressed to Bimu Yao, Lihui Bai and Wei Lu. 

\end{document}